\documentstyle[11pt,epsfig]{article}
\begin{document}

\title{Strong Decays of the Radial Excited States $B(2S)$ and $D(2S)$}
\vspace{10mm}

\author{Jin-Mei Zhang\footnote{jinmei\_zhang@tom.com}~~and
Guo-Li Wang\footnote{gl\_wang@hit.edu.cn} \\
\\
{\it \small  Department of Physics, Harbin Institute of
Technology, Harbin 150001, China} }
\date{}
\maketitle

\baselineskip=24pt

\begin{quotation}

\vspace*{1.5cm}
\begin{center}
  \begin{bf}
  ABSTRACT
  \end{bf}
\end{center}

\vspace*{0.5cm} \noindent The strong OZI allowed decays of the
first radial excited states $B(2S)$ and $D(2S)$ are studied in the
instantaneous Bethe-Salpeter method, and by using these OZI
allowed channels we estimate the full decay widths:
$\Gamma_{B^0(2S)}=24.4$ MeV, $\Gamma_{B^+(2S)}=23.7$ MeV,
 $\Gamma_{D^0(2S)}=11.3$ MeV and $\Gamma_{D^+(2S)}=11.9$ MeV.
 We also predict the masses of them:
$M_{B^0(2S)}=5.777$ GeV, $M_{B^+(2S)}=5.774$ GeV,
$M_{D^0(2S)}=2.390$ GeV and $M_{D^+(2S)}=2.393$ GeV.

\end{quotation}

\newpage
\setcounter{page}{1}

In the past few years, there are many new states observed in
experiments. Among them, the new states $D^{*}_{s0}(2317)$,
$D_{s1}(2460)$ \cite{2317}, $B_{s1}(5830)$ and $B_{s2}(5840)$
\cite{5830} are orbitally excited states, which are also called
$P$ wave states. So far, great progress has been made on the
physics of orbital excited states $D^{*}_{s0}(2317)$ and
$D_{s1}(2460)$ \cite{2317re}, and there are already exist some
investigations of $B_{s1}(5830)$ and $B_{s2}(5840)$ \cite{5840re}.
Around the energy of these hadrons, according to constitute quark
model, there may be the radial excited $S$ wave states $B(2S)$ and
$D(2S)$.  But due to their absence, the experimental and
theoretical studies for
 the radial excited 2$S$ states $B(2S)$ and $D(2S)$ are still
missing in the literature.

We know that the first radial excited 2$S$ state has a node
structure in its wave function, which means relativistic
correction of 2$S$ state is much larger than the one of
corresponding basic state, even the 2$S$ state is a heavy meson,
so to consider the physics of radial excited state a relativistic
method is needed. Bethe-Salpeter equation \cite{E51} and its
instantaneous one, Salpeter equation \cite{E52}, are famous
relativistic methods to describe the dynamics of a bound state. In
a previous letter \cite{CG}, we have solved the full Salpeter
equations for pseudoscalar mesons, the masses of first radial
excited 2$S$ states are obtained, they are $M_{B^0(2S)}=5.777$
GeV, $M_{B^+(2S)}=5.774$ GeV, $M_{D^0(2S)}=2.390$ GeV and
$M_{D^+(2S)}=2.393$ GeV.

The mass of $B(2S)$ is 310 MeV higher than the threshold of mass
scale of $B^*\pi$, but lower than the threshold of $B_s^*K$, and
the mass of $D(2S)$ is 240 MeV higher than the threshold of mass
scale of $D^*\pi$, but lower than the threshold of $D_s^*K$, so
the strong decays $B(2S)\rightarrow B^*+\pi$ and $D(2S)\rightarrow
D^*+\pi$ are OZI allowed strong decays, and they are dominate
decay channels of $B(2S)$ and $D(2S)$, respectively. In this
letter, we calculate the strong decay widths of $B(2S)\rightarrow
B^*+\pi$ and $D(2S)\rightarrow D^*+\pi$ in the framework of
Bethe-Salpeter method.

Since one of the final state is $\pi$ meson in the OZI allowed
$B(2S)$ or $D(2S)$ strong decay, we use the reduction formula, PCAC
relation and low energy theorem, so for the strong decays
(considering the $B^0(2S)\rightarrow B^{*+}\pi^-$ as an example)
shown in Fig. 1, the transition matrix element can be written as
\cite{W1}:
\begin{equation}
T=\frac{P_{f_2}^\mu}{f_{P_{f_2}}}\langle
B^{*+}(P_{f_1})|\bar{u}\gamma_\mu\gamma_5d|B^0(P)\rangle,
\end{equation}
where $P$, $P_{f_1}$ and $P_{f_2}$ are the momenta of the initial
state $B^0(2S)$, final states $B^{*+}$ and $\pi^-$, respectively,
and $f_{P_{f_2}}$ is the decay constant of $\pi^-$ meson.

To evaluate Eq. (1), we need to calculate the hadron matrix element
 $\langle
B^{*+}(P_{f_1})|\bar{u}\gamma_\mu\gamma_5d|B^0(P)\rangle$. It is
well known that the Mandelstam formalism \cite{mandelstam} is one
of proper approaches to compute the hadron matrix elements
sandwiched by the Bethe-Salpeter or Salpeter wave functions of two
bound-state. With the help of this method, in leading order, the
hadron matrix elements in the center of mass system of initial
meson can be written as \cite{W1,Chang}:
\begin{eqnarray}\label{a08}
\langle
B^{*+}(P_{f_1})|\bar{u}\gamma_\mu\gamma_5d|B^0_{2S}(P)\rangle=
\int\frac{d{\vec{q}}}{(2\pi)^3}Tr\left[
\bar{\varphi}^{++}_{_{P_{f_1}}}(\vec{q^\prime})\gamma_{\mu}\gamma_5
{\varphi}^{++}_{_P}({\vec{q}})\frac{\not\!P}{M} \right].
\end{eqnarray}
\begin{figure}
 \centering
\includegraphics[width=4.0in]{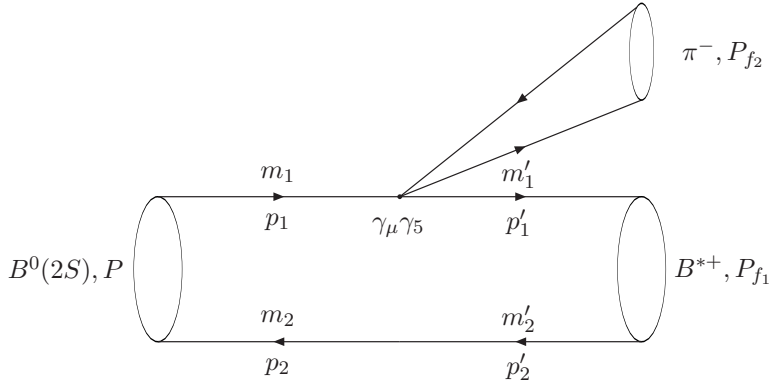}
\caption{\label{fig:epsart} Feynman diagram corresponding to the
strong decays $B^0(2S)\rightarrow B^{*+}\pi^-$.}
\end{figure}
where $\vec{q}$ is the relative three-momentum of the
quark-anti-quark in the initial meson $B^0(2S)$ and
$\vec{q^\prime}={\vec{q}}+\frac{m_2^\prime}{m_1^\prime+m_2^\prime}{\vec{r}}$,
$M$ is the mass of $B^0(2S)$, ${\vec{r}}$ is the three dimensional
momentum of the final meson $B^{*+}$, ${\varphi}^{++}_{_P}$ is the
positive energy B.S. wave function for the relevant mesons and
$\bar{\varphi}^{++}_{_{P_{f_1}}}=\gamma_0({\varphi}^{++}_{_{P_{f_1}}})^+\gamma_0$.

For the initial state pseudoscalar meson $B^0(2S)$ ($J^P=0^-$),
the positive energy wave function takes the general form
\cite{CG}:
\begin{eqnarray}\label{a06}
\varphi_{0^-}^{++}(\vec{q})&=&\frac{M}{2}\left\{\left[f_1(\vec{q})+f_2(\vec{q})
\frac{m_1+m_2}{\omega_1+\omega_2}\right]
\left[\frac{\omega_1+\omega_2}{m_1+m_2}+\frac{\not\!{P}}{M}
-\frac{\not\!{q_{_\bot}}(m_1-m_2)}{m_2\omega_1+m_1\omega_2}\right]\right.\nonumber\\
&+&\left.\frac{\not\!{q_{_\bot}}\not\!P(\omega_1+\omega_2)}{M(m_2\omega_1+m_1\omega_2)}\right\}\gamma_5.
\end{eqnarray}
where $q_{_\bot}=(0,\vec{q})$ and $\omega_i=\sqrt{m_i^2+\vec{q}^2}$,
$f_i(\vec{q})$ are eigenvalue wave functions which can be obtained
by solving the full $0^-$ state Salpeter equations. For the final
state vector meson $B^{*+}$ ($J^P=1^-$), the positive energy wave
function takes the general form \cite{05}:
\begin{eqnarray}
\varphi_{1^-}^{++}(\vec{q^\prime})&=&\frac{1}{2}
\left[A\not\!\epsilon_{_\bot}^{\prime\lambda}+B\not\!\epsilon_{_\bot}^{\prime\lambda}\not\!{P_{_{f_1}}}
+C(\not\!{q_{_\bot}^\prime}{\not\!\epsilon}_{_\bot}^{\prime\lambda}-q_{_\bot}^\prime\cdot\epsilon_{_\bot}^{\prime\lambda})
+D(\not\!{P_{_{f_1}}}\not\!\epsilon_{_\bot}^{\prime\lambda}\not\!{q_{_\bot}^\prime}
-\not\!{P_{_{f_1}}}q_{_\bot}^\prime\cdot\epsilon_{_\bot}^{\prime\lambda})\right.\nonumber\\
&+&\left.q_{_\bot}^\prime\cdot\epsilon_{_\bot}^{\prime\lambda}(E+F\not\!{P_{_{f_1}}}
+G\not\!{q_{_\bot}^\prime}+H\not\!{P_{_{f_1}}}\not\!{q_{_\bot}^\prime})\right],
\end{eqnarray}
where $\epsilon$ is the polarization vector of meson, and $A,\ B,\
C,\ D,\ E,\ F,\ G,\ H$ are defined as:
\begin{eqnarray}
A&=&M^\prime\left[f_5(\vec{q^\prime})-f_6(\vec{q^\prime})
\frac{\omega_1^\prime+\omega_2^\prime}{m_1^\prime+m_2^\prime}\right],\nonumber\\
B&=&\left[f_6(\vec{q^\prime})-f_5(\vec{q^\prime})
\frac{m_1^\prime+m_2^\prime}{\omega_1^\prime+\omega_2^\prime}\right],\nonumber\\
C&=&\frac{M^\prime(\omega_2^\prime-\omega_1^\prime)}{m_2^\prime\omega_1^\prime+m_1^\prime\omega_2^\prime}\left[f_5(\vec{q^\prime})
-f_6(\vec{q^\prime})\frac{\omega_1^\prime+\omega_2^\prime}{m_1^\prime+m_2^\prime}\right],\nonumber\\
D&=&\frac{\omega_1^\prime+\omega_2^\prime}{\omega_1^\prime\omega_2^\prime+m_1^\prime
m_2^\prime+\vec{q^\prime}^2}\left[f_5(\vec{q^\prime})
-f_6(\vec{q^\prime})\frac{\omega_1^\prime+\omega_2^\prime}{m_1^\prime+m_2^\prime}\right],\nonumber\\
E&=&\frac{m_1^\prime+m_2^\prime}{M^\prime(\omega_1^\prime\omega_2^\prime+m_1^\prime
m_2^\prime-\vec{q^\prime}^2)}
\left\{M^{\prime2}\left[f_5(\vec{q^\prime})-f_6(\vec{q^\prime})
\frac{m_1^\prime+m_2^\prime}{\omega_1^\prime+\omega_2^\prime}\right]
-\vec{q^\prime}^2\left[f_3(\vec{q^\prime})+f_4(\vec{q^\prime})
\frac{m_1^\prime+m_2^\prime}{\omega_1^\prime+\omega_2^\prime}\right]\right\},\nonumber\\
F&=&\frac{\omega_1^\prime-\omega_2^\prime}{M^{\prime2}(\omega_1^\prime\omega_2^\prime+m_1^\prime
m_2^\prime-\vec{q^\prime}^2)}
\left\{M^{\prime2}\left[f_5(\vec{q^\prime})-f_6(\vec{q^\prime})
\frac{m_1^\prime+m_2^\prime}{\omega_1^\prime+\omega_2^\prime}\right]
-\vec{q^\prime}^2\left[f_3(\vec{q^\prime})+f_4(\vec{q^\prime})
\frac{m_1^\prime+m_2^\prime}{\omega_1^\prime+\omega_2^\prime}\right]\right\},\nonumber\\
G&=&\left\{\frac{1}{M^\prime}\left[f_3(\vec{q^\prime})+f_4(\vec{q^\prime})
\frac{m_1^\prime+m_2^\prime}{\omega_1^\prime+\omega_2^\prime}\right]
-\frac{2f_6(\vec{q^\prime})M^\prime}{m_2^\prime\omega_1^\prime+m_1^\prime\omega_2^\prime}\right\},\nonumber\\
H&=&\frac{1}{M^{\prime2}}\left\{\left[f_3(\vec{q^\prime})
\frac{\omega_1^\prime+\omega_2^\prime}{m_1^\prime+m_2^\prime}+f_4(\vec{q^\prime})\right]
-2f_5(\vec{q^\prime})
\frac{M^{\prime2}(\omega_1^\prime+\omega_2^\prime)}{(m_1^\prime+m_2^\prime)(\omega_1^\prime\omega_2^\prime+m_1^\prime
m_2^\prime+\vec{q^\prime}^2)}\right\}.
\end{eqnarray}
where $M^{\prime}$ is the mass of $B^{*+}$, eigenvalue wave
functions $f_i(\vec{q^\prime})$ can be obtained by solving the full
$1^-$ state Salpeter equations.

In calculation of transition matrix element and solving the full
Salpeter equation, there are some parameters have to be fixed, the
input parameters are chosen as follows \cite{CG}: $m_b=5.224$
GeV,\ $m_c=1.7553$ GeV,\ $m_d=0.311$ GeV,\ $m_u=0.307$ GeV. The
values of the decay constants we use in this letter are
$f_{\pi^{\pm}}=0.1307$ GeV, $f_{\pi^0}=0.13$ GeV \cite{Eidelman}.
With the parameters, the masses of the radial excited 2$S$ states
are present: $M_{B^0(2S)}=5.777$ GeV, $M_{B^+(2S)}=5.774$ GeV,
$M_{D^0(2S)}=2.390$ GeV and $M_{D^+(2S)}=2.393$ GeV. The numerical
strong decay widths of $B(2S)$ and $D(2S)$ mesons are shown in
Table 1.

In our results only the $1^-0^-$ final states are calculated
($B^*\pi$ and $D^*\pi$), in our estimate of mass spectra, there are
no other OZI allowed strong decay channels. For example, from the
analysis of quantum number, there may be the decay channels with $P$
wave in the final states, for example, the final state can be
$0^+0^-$ $B(1P)\pi$ states, but due to our estimate the mass of
lightest $P$ wave $0^+$ state $m_{B(1P)}=5.665$ GeV \cite{photon},
which is larger than the threshold of $B(2S)$ (the same results for
$D(2S)$ cases), so there is no phase space for this channel, if
later experimental discovery of mass of this state is lower than
theoretical estimate like happened to $D_{s0}(2317)$, which has been
hoped much higher than $2317$ MeV, this channel become a OZI allowed
one, but because the phase space is very small, and it is a $P$
wave, the transition decay width should be smaller than the case
when it is $S$ wave, so we can ignore the contributions of these
channels and other electroweak channels, and we use these OZI
allowed decay widths to estimate the full decay width of this 2$S$
state.

\begin{table*}
\setlength{\tabcolsep}{0.8cm} \caption{\small The strong decay
widths of the 2$S$ state $B$ and $D$ mesons.} \doublerulesep2pt
\begin{tabular}{||c|c|c|c||}
\hline \hline Mode&$\Gamma$ (MeV)&Mode&$\Gamma$ (MeV)
\\ \hline
\ \ $B^0(2S)\rightarrow B^{*+}\pi^-$&12.3&\ \ $D^0(2S)\rightarrow
D^{*+}\pi^-$&5.48
\\ \hline
$B^0(2S)\rightarrow B^{*0}\pi^0$&12.1&$D^0(2S)\rightarrow
D^{*0}\pi^0$&5.85
\\ \hline
$B^+(2S)\rightarrow B^{*0}\pi^+$&11.7&$D^+(2S)\rightarrow
D^{*0}\pi^+$&6.05
\\ \hline
$B^+(2S)\rightarrow B^{*+}\pi^0$&12.0&$D^+(2S)\rightarrow
D^{*+}\pi^0$&5.80
\\ \hline\hline
\end{tabular}
\end{table*}

This work was supported in part by the National Natural Science
Foundation of China (NSFC) under Grant No. 10875032, and in part
by SRF for ROCS, SEM.

\end{document}